\documentclass{emulateapj}
\usepackage{apjfonts}
\usepackage{rotating}
\usepackage{enumerate}

\newcommand{\pivec}{\mbox{\boldmath $\pi$}}

\renewcommand{\thefootnote}{\ifcase\value{footnote}\or{$\dagger$} \or{$\ddagger$} \or($\infty$)\fi}

\lefthead{SHIN ET AL.} 
\righthead{Constraints on Additional Planets in Known Planetary Systems
}

\begin{document}
\title{
Constraint on Additional Planets in Planetary Systems Discovered 
through the Channel of High-magnification Gravitational Microlensing Events
}

\author{
I.-G. Shin,
C. Han$^{\dagger}$, 
J.-Y. Choi,
K.-H. Hwang,
Y. K. Jung,
H. Park\\
%
}

\bigskip\bigskip
\affil{Department of Physics, Institute for Astrophysics, Chungbuk National University, Cheongju 371-763, Republic of Korea}
\affil{$^{\dagger}$Corresponding author} 

\begin{abstract}
High-magnification gravitational microlensing events provide an important channel of detecting planetary systems with 
multiple giants located at their birth places. In order to investigate the potential existence of additional planets, 
we reanalyze the light curves of the eight high-magnification microlensing events for each of which a single planet was 
previously detected. The analyzed events include OGLE-2005-BLG-071, OGLE-2005-BLG-169, MOA-2007-BLG-400, MOA-2008-BLG-310, 
MOA-2009-BLG-319, MOA-2009-BLG-387, MOA-2010-BLG-477, and MOA-2011-BLG-293. We find that including an additional planet 
improves fits with $\Delta\chi^2 < 80$ for seven out of eight analyzed events. For MOA-2009-BLG-319, the improvement is 
relatively big with $\Delta\chi^2 \sim 143$. From inspection of the fits, we find that the improvement of the fits is 
attributed to systematics in data. Although no clear evidence of additional planets is found, it is still possible to 
constrain the existence of additional planets in the parameter space. For this purpose, we construct exclusion diagrams 
showing the confidence levels excluding the existence of an additional planet as a function of its separation and mass 
ratio. We also present the exclusion ranges of additional planets with 90\% confidence level for Jupiter, Saturn, and 
Uranus-mass planets.
\end{abstract}


\section{Introduction}
Since the first discoveries \citep{wolszczan92, mayor95}, the number of known extra-solar planets has been rapidly 
increasing and it reaches $\sim2000$ \citep[http://exoplanet.eu:][]{schneider11}. A significant fraction of these 
planets are members of multi-planet systems \citep{rowe14, lissauer12}, which were mostly discovered by the 
radial-velocity and transit methods. 

Multi-planet systems can also be detected by using the microlensing method. Gravitational microlensing occurs by 
the chance alignment along the line of sight toward a background star (source) and a foreground object (lens), 
resulting in the magnification of the source brightness due to the bending of light by the gravity of the lens. 
When a lensing object is a star accompanied by a planet, the planet causes astigmatism in the bending of light 
and induces formation of caustics which refers to the envelope of light rays on the source plane at which the 
magnification of a point source becomes infinity. A planetary lensing signal occurs when the source approaches 
caustics and it appears as a short term anomaly superposed on the smooth and symmetric lensing light curve of 
the host star \citep{mao91, gould92b}. If a planetary system is composed of multiple planets, 
each planet induces its own caustics and the multiplicity of the planets can be identified if the source trajectory 
passes multiple caustics induced by the individual planets \citep{han01, han05}. The sensitivity to multiple planets 
for general microlensing events is low because planet-induced caustics are small and thus the geometrical probability 
for the source to pass multiple caustics is low. However, the sensitivity is high for a subset of lensing events where 
the brightness of a source star is greatly magnified \citep{gaudi98}. This is because one of the caustics induced by each 
planet is always located very close to the host star around which the source of a high-magnification event 
passes by \citep{griest98}. Indeed, two multi-planet systems identified by microlensing \citep{gaudi08, han13} were 
discovered through this channel. 

Microlensing discoveries of multi-planet systems are important for better understanding of planet-formation mechanism. 
According to the core accretion theory \citep{ida04}, giant planets are believed to form in group in the region beyond 
the snow line where the lower temperature in this region makes many more solid grains available for accretion into 
planetesimals. However, most known planets in multiple systems reside well within the snow 
line. These close-in planets are believed to have migrated from the outer region where they were formed 
\citep{dangelo08, lubow11}.  Detecting planets at their birth places by using the transit and the radial-velocity methods 
is difficult due to their low sensitivity to wide-separation planets. By contrast, microlensing is sensitive to planets 
located around and beyond the snow line and thus provides a channel to study planets where they formed. 
 
In this work, we reanalyze the light curves of the eight high-magnification microlensing events for each of which a 
single planet was previously detected. From the analyses, we investigate the existence of additional planets. 

\begin{deluxetable*}{lrrrl}
\tablecaption{List of sample events and the properties of the previously dicovered planets\label{table:one}}
\tablewidth{0pt}
\tablehead{
\multicolumn{1}{c}{Event} &
\multicolumn{1}{c}{$A_{\rm max}$} &
\multicolumn{2}{r}{Previously dicovered planet} &
\multicolumn{1}{c}{Reference} \\
\multicolumn{1}{c}{} &
\multicolumn{1}{c}{} &
\multicolumn{1}{c}{Mass} &
\multicolumn{1}{c}{Separation (AU)} &
}
\startdata
OGLE-2005-BLG-071  &  59 &  3.8$^{+0.4}_{-0.4}$ $M_{\rm J}$   &  3.6$^{+0.2}_{-0.2}$                         & \citet{udalski05}, \citet{dong09a} \\
OGLE-2005-BLG-169  & 880 &   13$^{+6}_{-8}$ $M_{\oplus}$      &  2.7$^{+1.7}_{-1.4}$                         & \citet{gould06}                    \\
MOA-2007-BLG-400   & 629 & 0.83$^{+0.49}_{-0.31}$ $M_{\rm J}$ & 0.72$^{+0.38}_{-0.16}$ / 6.5$^{+3.2}_{-1.2}$ & \citet{dong09b}                    \\
MOA-2008-BLG-310   & 366 &   74$^{+17}_{-17}$ $M_{\oplus}$    & 1.25$^{+0.10}_{-0.10}$                       & \citet{janczak10}                  \\
MOA-2009-BLG-319   & 208 &   50$^{+44}_{-24}$ $M_{\oplus}$    &  2.4$^{+1.2}_{-0.6}$                         & \citet{miyake11}                   \\
MOA-2009-BLG-387   &  49 &  2.6$^{+4.1}_{-1.6}$ $M_{\rm J}$   &  1.8$^{+0.9}_{-0.7}$                         & \citet{batista11}                  \\
MOA-2010-BLG-477   & 599 &  1.5$^{+0.8}_{-0.3}$ $M_{\rm J}$   &  2.0$^{+3.0}_{-1.0}$                         & \citet{bachelet12}                 \\
MOA-2011-BLG-293   & 418 &  2.4$^{+1.5}_{-0.9}$ $M_{\rm J}$   &  1.0$^{+0.1}_{-0.1}$ / 3.4$^{+0.4}_{-0.4}$   & \citet{yee12}                      
\enddata
\end{deluxetable*}

\section{Sample}
In published articles, there exist 29 planetary microlensing events from which 31 planets were found. Among them, events 
for our analyses are selected based on the following criterion. First, we choose planetary events with high magnifications 
due to their high sensitivity to multiple planets. With an adopted threshold magnification $A_{\rm th}\sim50$, there exist 
15 events. Among the planetary systems found from these events, two are already known to be in multi-planet systems\footnote{
The first known multiple microlensing planetary system is OGLE-2006-BLG-109L, where a host star with a mass 
$\sim0.5$ $M_{\odot}$ is orbited by two planets with masses $\sim0.71$ $M_{\rm J}$ and $\sim0.27$ $M_{\rm J}$ and 
separations from the host $\sim2.3$ AU and $\sim4.6$ AU, respectively \citep{gaudi08, bennett10}. The second system is OGLE-2012-BLG-0026L, 
where a host star with a mass $\sim0.82$ $M_{\odot}$ is orbited by two planets with masses $\sim0.11$ $M_{\rm J}$ 
and $\sim0.67$ $M_{\rm J}$ and separations $\sim3.8$ AU and $\sim4.6$ AU, respectively \citep{han13}. } and another 
two are in stellar binary systems.\footnote{
OGLE-2013-BLG-0341Lb is the first microlensing planet in a binary stellar system, where a terrestrial planet 
($\sim2$ $M_{\oplus}$) lies at $\sim0.8$ AU from its host with a mass $\sim$0.10--0.15 $M_{\odot}$ and the host 
itself orbits another star with a mass $\sim0.15$ $M_{\odot}$ and a separation $\sim$10--15 AU \citep{gould14}. 
The second microlensing planet in binary is OGLE-2008-BLG-092Lb, where a $\sim4$ $M_{\oplus}$ planet orbits a 
$\sim0.7$ $M_{\odot}$ star at $\sim18$ AU. The host has a low-mass companion with $\sim0.15$ $M_{\odot}$ 
\citep{poleski14}. } We exclude the events with 
already-known multi-planet systems and circum-stellar binary planets from our analysis. Second, we choose events with good 
coverage of the peak region. This is because signatures of multiple planets for high-magnification events show up near the 
peak of a light curve and thus good coverage of the peak region is essential to constrain additional planets. Third, we 
confine analysis to events for which there is no ambiguity in the interpretation of the known planetary signals. Applying 
these criteria leave 8 events including OGLE-2005-BLG-071, OGLE-2005-BLG-169, MOA-2007-BLG-400, MOA-2008-BLG-310, 
MOA-2009-BLG-387, MOA-2010-BLG-477, and MOA-2011-BLG-293.

\begin{figure*}[ht]
\epsscale{0.95}
\plotone{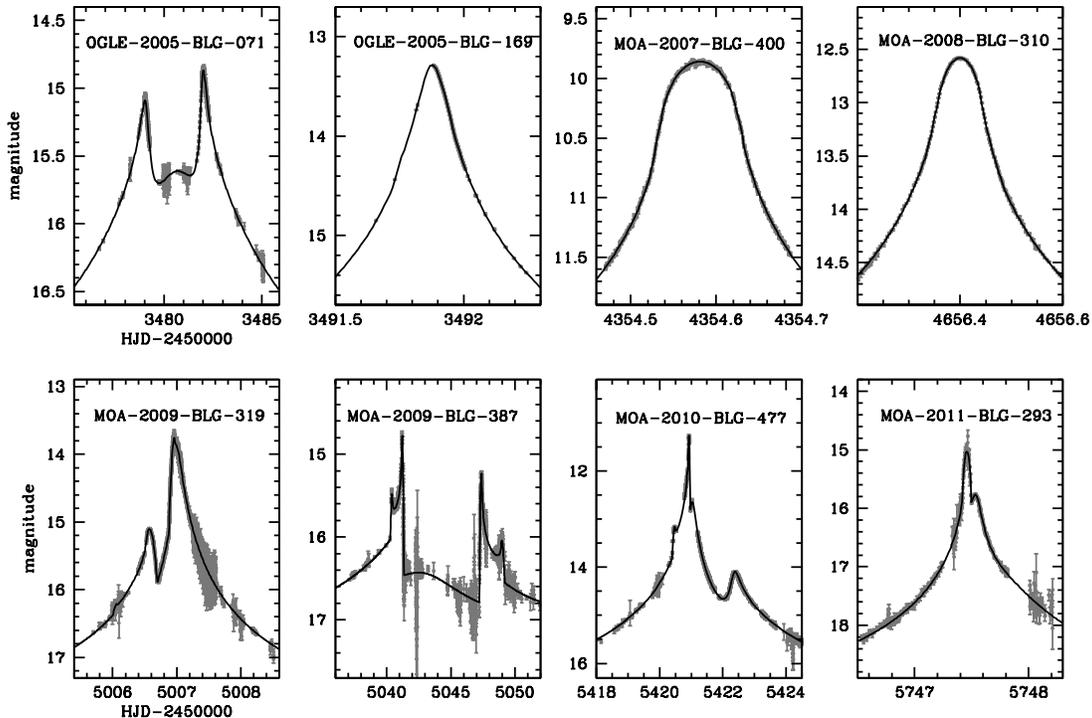}
\caption{\label{fig:one}
Light curves of planetary microlensing events reanalyzed in this work. Grey points with error bars 
are the data used in our analysis. Also presented are the model light curves base on single-planet 
models.
}\end{figure*}

In Table \ref{table:one}, we list the lensing events that we analyzed. Also presented are the peak magnifications 
$A_{\rm max}$ of the individual events,the physical parameters (mass and separation from the host) of the previously 
discovered planets, and the references of the original analyses. In Figure \ref{fig:one}, we present the light curves 
of the events. We note that the same data sets as those used in the original analysis are used in our analysis for 
the consistency of results.

\section{Analysis Method}
In order to investigate the existence of an additional planet for a lensing event with an already known planet, it is 
required to conduct triple-lens modeling. For a triple-lens system, the image positions for given locations of the lens 
and source are found by solving the equation of lens mapping (lens equation) that is expressed as  
\begin{equation}
\zeta=z-\sum\limits_{k=1}^3 { \epsilon_{k} \over {\bar{z}-\bar{z}_{{\rm L},k}} }.
\end{equation}
Here $k=$1, 2, and 3 denote the individual lens components, $\zeta=\zeta+i\eta$, 
$z_{{\rm L},k}=x_{{\rm L},k} + iy_{{\rm L},k}$ and $z=x+iy$ are the complex notations of the source, lens, and image 
positions, respectively, $\bar{z}$ denotes the complex conjugate of $z$, and $\epsilon_{k}=M_{k}/M_{\rm tot}$ is the 
mass fraction of each lens component \citep{witt90}. Once the image locations are found, the lensing magnification is 
computed by
\begin{equation}
A=\sum {A_{i}}; ~~A_{i}= \left| \left( 1- \frac{\partial\zeta}{\partial\bar{z}} \frac{\partial\bar{\zeta}}{\partial\bar{z}} 
\right)^{-1}_{z=z_{i}} \right|,
\end{equation}
where $A_{i}$ is the magnification of each image.

Due to non-linearity, the lens equation for a triple lens system cannot be analytically solved. However, lensing 
magnifications can be numerically computed by using the method known as the inverse ray-shooting method 
\citep{kayser86, schneider86}. In this method, a large number of rays are uniformly shot from the image plane, 
bent by the lens equation, and land on the source plane. Then, lensing magnifications are computed as the ratio 
of the number density of rays on the source plane to that on the image plane. The lens equation can also be solved 
semi-analytically because the equation is expressed as a tenth-order polynomial in $z$ and the image positions 
and subsequent lensing magnifications are computed by solving the polynomial \citep{rhie02, song14}. The 
semi-analytic method has an advantage of fast computation but it is difficult to be used for computing lensing 
magnifications affected by finite-source effects. On the other hand, the numerical method can be used to compute 
finite-source magnifications but computation is slow.

Describing the light curve of a triple-lensing event requires ten basic parameters. The first three of these 
parameters describe how the source approaches the lens. They are the time of the closest source approach to 
a reference position of the lens system, $t_0$, the normalized projected separation between the source and the lens 
reference position at that moment, $u_0$, and the Einstein time scale, $t_{\rm E}$, that is defined as the time 
required for the source to cross the angular Einstein radius of the lens $\theta_{\rm E}$. Another six parameters 
characterize the lens system. They are the projected separations between the host and the individual planet 
companions, $s_1$ and $s_2$, and the mass ratios of the companions to the host, $q_1$ and $q_2$, the angle between 
the source trajectory and the axis connecting the host and the first planet, $\alpha$, and the orientation angle 
of the second planet with respect to the host-first planet axis, $\psi$. We note that length scales 
describing lensing phenomenon are usually expressed in units of $\theta_{\rm E}$ and $u_0$, $s_1$ and $s_2$ are 
normalized by $\theta_{\rm E}$. The last parameter is the angular size of the source $\theta_{\ast}$ in unit of 
the Einstein radius, $\rho_{\ast}=\theta_{\ast}/\theta_{\rm E}$ (normalized source radius). This parameter is 
needed to describe light curve deviations affected by finite-source effects.

Besides the basic lensing parameters, precise description of lensing light curves often requires additional 
parameters to account for higher-order effects. One such effect is caused by the positional change of an observer 
induced by the orbital motion of the Earth around the Sun. Considering this parallax effect \citep{gould92a} requires 
two additional parameters $\pi_{{\rm E},N}$ and $\pi_{{\rm E},E}$ that represent the two components of the lens parallax 
vector $\pivec_{\rm E}$ projected on the sky in the north and east equatorial coordinates, respectively. Similarly, 
lensing light curves can also be affected by the change of the lens position caused by the orbital motion 
\citep{shin11, shin12, skowron11, park13}.

With the lensing parameters, we conduct triple-lens modeling of the events that were previously analyzed based on binary-lens 
modeling. For all analyzed events, the anomaly is dominated by the signal of the already reported planet and thus the signal 
of the potential additional planet would be small. We, therefore, set the lensing parameters related to the known planet 
$(s_1, q_1)$ fixed as outlined in \citet{kubas08}. The existence of an additional planet is then inspected by checking whether 
the two-planet model improves the fit with respect to the single-planet model. Not knowing the characteristics of the second 
planet, we search for the second-planet parameters $s_2$ and $q_2$ by inspecting solutions in wide ranges of the parameter 
space spanning $-1\leq {\rm log} s_2 \leq 1$ and $-6\leq {\rm log} q_2 \leq 2$, respectively. For a given set of $(s_2, q_2)$, 
the other lensing parameters are searched for by minimizing $\chi^2$ in the parameter space using Markov Chain Monte Carlo 
(MCMC) algorithm.

For computing triple-lensing magnifications, we apply both the semi-analytic and numerical methods. For the region 
near the peak of the light curve, we use the numerical inverse ray-shooting method because this region is likely to 
be affected by finite-source effects during the source star's approach close to the planet-induced caustics. For 
other regions, we use the semi-analytic method for fast computation. Computing finite-source magnifications is further 
accelerated by applying the ``map making'' method \citep{dong06}. In this method, a map of rays for a given set of 
$(s_1, q_1)$ and $(s_2, q_2)$ is constructed based on the inverse ray shooting method and then it is used to produce 
many different light curves resulting from various source trajectories without producing extra maps. In computing 
finite-source magnifications, we consider limb-darkening variation of the source star surface brightness by modeling 
the surface brightness profile as $S_{\lambda} \propto 1 - \Gamma_{\lambda}(1-1.5\cos{\phi})$, where $\Gamma_{\lambda}$ 
is the limb-darkening coefficients corresponding to specific passband $\lambda$ and $\phi$ is the angle between the line 
of sight toward the source star and the normal to the source surface. We use the same limb-darkening coefficients that 
were used in the previous analyses where first planets were reported.

If higher-order effects were reported in the previous analyses, we also consider them. Parallax effects were reported for events 
OGLE-2005-BLG-071, MOA-2009-BLG-387, and MOA-2010-BLG-477. On the other hand, obvious effect of lens-orbital motion was reported 
for none of the events. We note that the effect of an additional planet is confined to the narrow region of the light curve peak 
while both parallax and lens-orbital effects influence the overall shape of a lensing light curve. Therefore, the existence of 
the second planet does not affect the lensing parameters of higher-order effects.

\section{Result}

\subsection{Existence of Additional Planets}

Table \ref{table:two} shows the result of the analysis presented as $\chi^2$ values of the binary and triple lens models. 
From the comparison of the $\chi^2$ values, it is found that including an additional planet improves fits with 
$\Delta\chi^2 < 80$ for seven out of the eight analyzed events. For MOA-2009-BLG-319, the improvement is relatively big 
with $\Delta\chi^2=142.7$.

\begin{deluxetable*}{lrrrccc}
\tablecaption{Comparision of binary and triple lens fits and exclusion range of a second planet\label{table:two}}
\tablewidth{0pt}
\tablehead{
\multicolumn{1}{c}{Event} &
\multicolumn{1}{c}{$\chi^2_{\rm bi}$} &
\multicolumn{1}{c}{$\chi^2_{\rm tri}$} &
\multicolumn{1}{c}{$\Delta\chi^2$} &
\multicolumn{3}{c}{Exclusion range (AU)} \\
\multicolumn{1}{c}{} &
\multicolumn{1}{c}{} &
\multicolumn{1}{c}{} &
\multicolumn{1}{c}{} &
\multicolumn{1}{c}{Jupiter} &
\multicolumn{1}{c}{Saturn} &
\multicolumn{1}{c}{Uranus} 
}
\startdata
OGLE-2005-BLG-071  & 1305.6  & 1283.7  &  21.9 & 1.5 --  5.1 &     --     &     --     \\
OGLE-2005-BLG-169  &  533.4  &  518.5  &  14.9 & 0.6 -- 11.3 & 1.0 -- 7.1 &     --     \\
MOA-2007-BLG-400   &  770.8  &  760.0  &  10.8 & 0.8 --  4.5 & 1.2 -- 2.6 &     --     \\
MOA-2008-BLG-310   & 3154.4  & 3076.4  &  78.0 & 0.7 --  2.0 & 1.1 -- 1.3 &     --     \\
MOA-2009-BLG-319   & 6180.2  & 6037.5  & 142.7 & 0.5 --  7.9 & 0.9 -- 5.0 & 1.6 -- 2.9 \\
MOA-2009-BLG-387   & 1298.4  & 1269.3  &  29.1 & 0.9 --  4.4 & 1.6 -- 2.5 &     --     \\
MOA-2010-BLG-477   & 4444.0  & 4394.4  &  49.6 & 0.8 -- 12.1 & 1.7 -- 5.8 &     --     \\
MOA-2011-BLG-293   & 5089.2  & 5069.9  &  19.3 & 1.0 --  3.5 &     --     &     --
\enddata
\tablecomments{
The $\chi^2$ difference represents $\Delta\chi^2=\chi^2_{\rm bi} - \chi^2_{\rm tri}$, 
where $\chi^2_{\rm bi}$ and $\chi^2_{\rm tri}$ are for the binary and triple 
lens models, respectively.
The exclusion ranges are estimated based on 90\% confidence level
}
\end{deluxetable*}

\begin{figure}[ht]
\epsscale{1.1}
\plotone{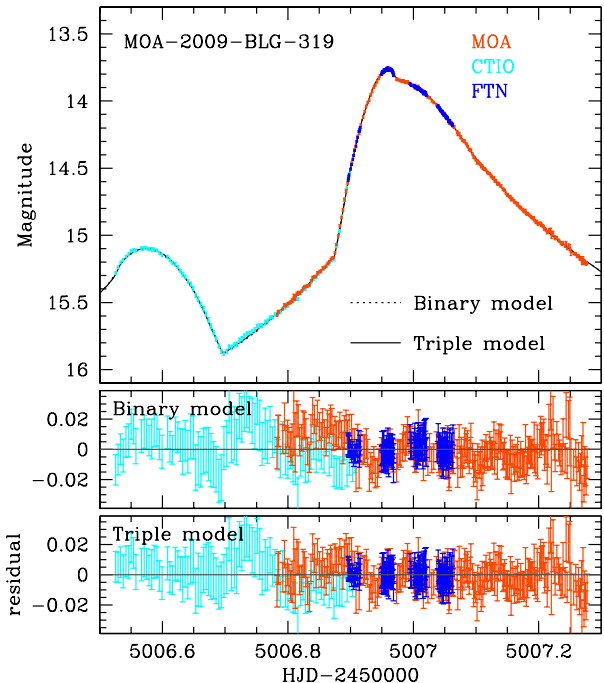}
\caption{\label{fig:two}
Comparison of the triple and binary models of MOA-2009-BLG-319. The lower two panels show 
the residuals from the individual models.
}\end{figure}

Although the improvement of the fit for the event MOA-2009-BLG-319 is formally significant, one cannot exclude the 
possibility of systematics in data, which is often masqueraded as planetary signals. In order to check the genuineness 
of the additional planetary signal, we inspect the fits from the triple and binary lens models. Figure \ref{fig:two} 
shows the peak portion of the light curve. We plot data sets from MOA, CTIO, and FTN observations for which most signal 
of the additional planet comes from: $\Delta\chi^2=78.4$ for MOA, $\Delta\chi^2=11.0$ for CTIO, and $\Delta\chi^2=35.5$ 
for FTN data.  Also plotted are the light curves of the triple and binary models. It is found that the two models cannot 
be distinguished within the resolution of the plot. To better show the difference between the two models, we also present 
the residuals from the individual models. We find that the improvement by the triple-lens fit is $\sim0.01$ mag level 
which is equivalent to the scatter of the data, implying that the signal can centainly be attributed to the systematics 
in data. We find similar results for other events. Therefore, we conclude that there is no case of firm detection of 
additional planets.

\subsection{Constraints on Additional Planets}
Although no clear evidence of additional planets is found, it is still possible to constrain the existence of additional 
planets around the lens. The constraint is expected to be strong because all analyzed events are highly magnified and 
thus efficiency to planets would be high in wide ranges of planet parameters. 

For the purpose of constraining additional planets, we construct 
\textit{exclusion diagrams} \citep{gaudi00, albrow00, kubas08, gould10, cassan12} showing the confidence levels excluding 
the existence of an additional planet as a function of its separation and mass ratio. To construct exclusion diagrams, we 
first estimate the detection efficiency of planets with respect to the planet parameters $s_2$ and $q_2$ according to the 
following procedure. 
\begin{enumerate}[(1)]
  \item We first choose a triple-lens configuration with $(s_2,q_2,\psi)$. For a given set of these parameters, 
        we determine the remaining parameters that best fit the observed light curve and then compute $\Delta\chi^2$ 
        between the triple and binary fits.
  \item We repeat the above process for many different orientation angles $\psi$ of the second planet.
  \item The detection efficiency for a given $(s_2,q_2)$ is estimated as the fraction of the angles $\psi$ that 
        produce significant deviations in the light curve. Then, the confidence level of excluding an additional 
        planet corresponds to the efficiency. 
\end{enumerate}
As a criterion for planet detections, we set a threshold value $\Delta\chi^2_{\rm th}$. \citet{gould10} suggested plausible 
values in the range $\Delta\chi^2_{\rm th}=350-700$ based on their experience in fitting lensing light curves for which 
planets were detected through the high-magnification channel. \citet{cassan12} suggested a similar threshold. In our analyses, 
we adopt a median value $\Delta\chi^2_{\rm th}=500$.

\begin{figure*}[ht]
\epsscale{0.80}
\plotone{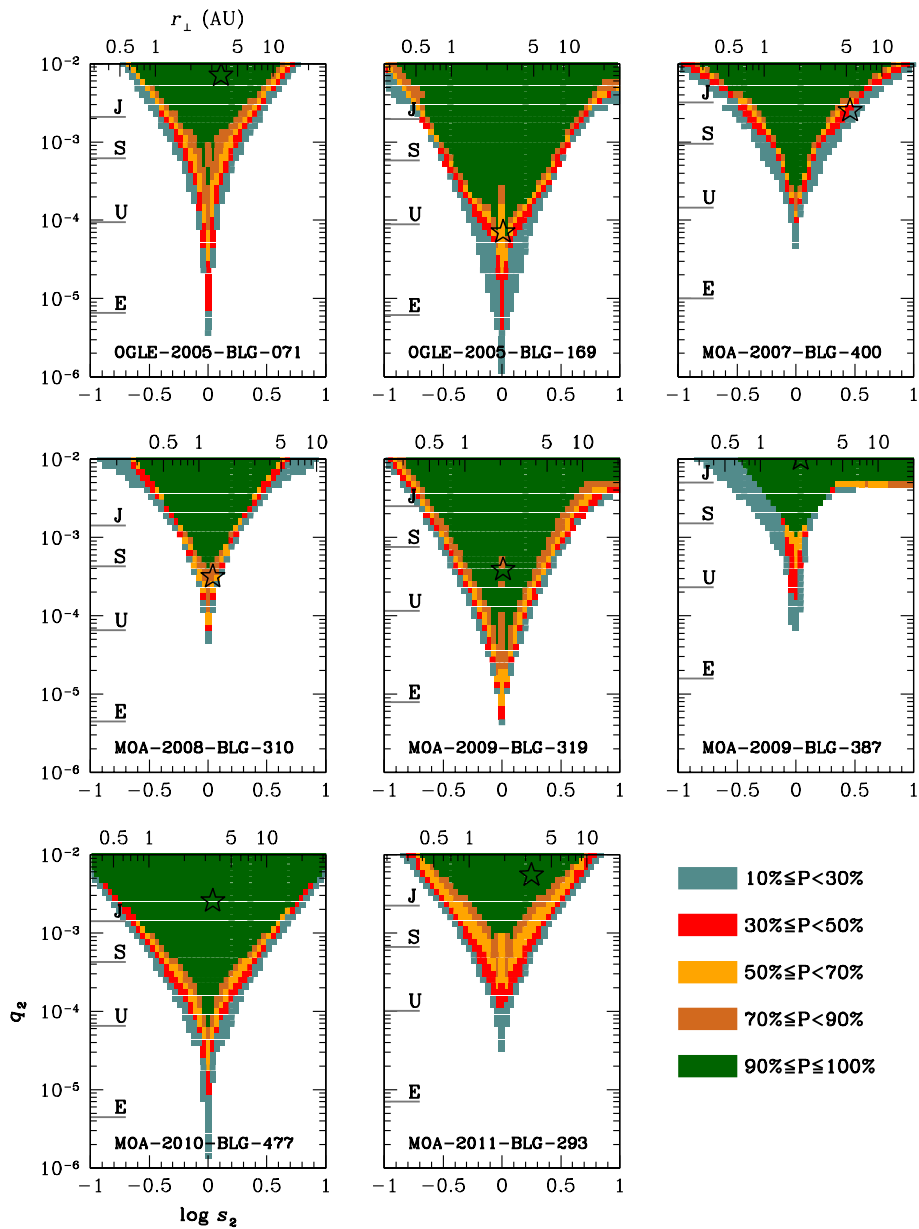}
\caption{\label{fig:three}
Exclusion diagrams showing the confidence levels excluding the existence of an additional planet 
with respect to the separation $s_2$ and the mass ratio $q_2$. For each panel, the upper $x$ label 
represents the planet-host separation in physical units. The ticks marked by ``J'', ``S'', ``U'', 
and ``E'' represent the planet masses corresponding to the Jupiter, Saturn, Uranus, and Earth of the 
Solar system. The star marks in the individual diagrams represent the planets detected from the 
previous analyses.
}\end{figure*}

Figure \ref{fig:three} shows the exclusion diagrams of the individual events. For each diagram, the regions 
with different confidence levels of excluding additional planets are marked in different colors. The star mark 
corresponds to the planet detected in the previous analysis.  The upper $x$ label represents the projected 
planet-host separation in a physical unit (AU). The ticks marked by ``J'', ``S'', ``U'', and ``E'' represent 
the planet masses corresponding to the Jupiter, Saturn, Uranus, and Earth of the Solar system, respectively. 
We note that the conversion from $(s_2,q_2)$ to the physical separation and mass of the planet is done based 
on the physical Einstein radius $r_{\rm E}$ and total mass of the lens system $M_{\rm tot}$ determined in the 
previous analyses, i.e., $r_{\bot}=sr_{\rm E}$ and $M_{{\rm p},2}=[q_2/(1+q_1+q_2)]M_{\rm tot}$.

In Table \ref{table:two}, we present the exclusion ranges of additional planets determinded with 90\% confidence 
level for Jupiter, Saturn, and Uranus-mass planets. It is found that the constraint on giant planets is considerably 
strong. On the other hand, the constraint on planets with masses less than Uranus is relatively weak. The weak 
constraint on low-mass planets is due to the small size of the planet-induced central caustic which decreases rapidly 
with the decrease of the planet mass \citep{chung05}. In addition to planet masses, the sensitivity depends on various 
other factors. Some of these factors are related to observation, including the photometric precision, cadence, and 
completeness of the planetary signal coverage, etc. Other factors are related to the intrinsic properties of planetary 
events, including the peak magnifications, severeness of finite-source effects, etc.

For OGLE-2005-BLG-169, MOA-2009-BLG-319, and MOA-2010-BLG-477, the constraint is strong. These events have a common 
property of high peak magnifications with $A_{\rm max}=880$, $208$, and $599$, respectively.

By contrast, the constraint is relatively weak for OGLE-2005-BLG-071 $(A_{\rm max}=59)$ and MOA-2009-BLG-387 
$(A_{\rm max}=49)$ because the peak magnifications are low. For planets either with low masses on located away from 
the Einstein ring of the host, the central caustic is small and thus signals of low-mass planets can be detected only 
for very high-magnification events where the source trajectories approach close to the caustic.

For MOA-2007-BLG-400 and MOA-2008-BLG-310, the peak magnifications are high: $A_{\rm max}=629$ and $A_{\rm max}=366$, 
respectively. Nevertheless, the sensitivity is relatively low. We find the main cause of the low sensitivity for these 
events is severe finite-source effects, which wash out planetary signals, especially for low-mass planets \citep{bennett96}.

For MOA-2011-BLG-293, on the other hand, the sensitivity is not very high although the event was highly magnified 
$(A_{\rm max}=418)$ and experienced little finite-source effects. We find that the low sensitivity is mainly due to 
low photometric precision caused by the source faintness.

\section{Conclusion}
In order to investigate the potential existence of additional planets, we reanalyzed high-magnification microlensing events for each 
of which a single planet had been previously detected. We found that introducing additional planets improves fits but the levels of 
improvement are not big enough to firmly identify additional planets. Although no clear evidence of additional planets was found, we 
could constrain the existence of additional planets in the parameter space. For this purpose, we presented exclusion diagrams showing 
the confidence levels excluding the existence of an additional planet as a function of its separation and mass ratio. We also presented 
the exclusion ranges of additional planets with 90\% confidence level for Jupiter, Saturn, and Uranus-mass planets.


\acknowledgments 
This work was supported by the research grant of Chungbuk National University in 2012.

\clearpage

\end{document}